\RequirePackage{fix-cm}
\let\accentvec\vec
\documentclass[twocolumn]{svjour3}
\smartqed  

\let\vec\accentvec
\usepackage{graphicx}
\usepackage[latin1]{inputenc}
\usepackage{amsmath,amsfonts,amssymb}
\usepackage[english]{babel}
\usepackage[T1]{fontenc}
\usepackage{textcomp}
\usepackage[subnum]{cases}
\journalname{Eur. Phys. J. C}
\date{November 2017}
\begin{document}
\title{Generalization of Einstein's gravitational field equations}
\author{Frédéric Moulin}
\institute{Département de Physique, Ecole Normale Supérieure Paris-Saclay, 61 av du président Wilson, F-94235 Cachan, France\\
\email{frederic.moulin@ens-paris-saclay.fr}}
\maketitle
\begin{abstract}
The Riemann tensor is the cornerstone of general relativity, but as everyone knows it does not appear explicitly in Einstein's equation of gravitation. This suggests that the latter may not be the most general equation. We propose here for the first time, following a rigorous mathematical treatment based on the variational principle, that there exists a generalized 4-index gravitational field equation containing the Riemann curvature tensor linearly, and thus the Weyl tensor as well. We show that this equation, written in $n$ dimensions, contains the energy-momentum tensor for matter and also that of the gravitational field itself. This new 4-index equation remains completely within the framework of general relativity and emerges as a natural generalization of the familiar 2-index Einstein equation. Due to the presence of the Weyl tensor, we show that this equation contains much more information, which fully justifies the use of a fourth-order theory. 
%\PACS{04.20.Fy \and 04.50.+h \and 04.50.Kd}
%
\end{abstract}
\section{Introduction}
Faisons maintenant varier l'action par rapport à la métrique:  \\
The Riemann curvature tensor of general relativity $R_{ijkl}$ can be split into the Weyl conformal tensor $C_{ijkl}$, and parts which involve only the Ricci tensor $R_{jl}$ and the curvature scalar $R$. Because of the properties of the Weyl tensor, its contraction vanishes, $g^{ik}C_{ijkl}=0$, and thus the information it contains (namely the information about the gravitational field in vacuum) is not present in the famous Einstein equation.\\
The aim of this paper is to find a generalized gravitational field equation explicitly containing the Riemann curvature tensor linearly. For this, we have implemented a rigorous mathematical treatment with a classical variational principle using a generalized Lagrangian containing $R_{ijkl}$, $R_{jl}$ and $R$. \\
The paper is organized as follows: \\
In section 2 we generalize the Einstein-Hilbert Lagrangian of general relativity by introducing new scalars constructed from $R_{ijkl}$ and $R_{jl}$. \\
In sections 3 and 4 we apply the principle of least action to this Lagrangian and obtain the generalized 4-index Einstein equation written with the total energy-momentum tensor, $T_{ijkl}=T^{(M)}_{ijkl}+T^{(F)}_{ijkl}$, where $T^{(M)}_{ijkl}$ is the energy-momentum tensor for matter, and $T^{(F)}_{ijkl}$ the energy-momentum tensor for the gravitational field itself. We show that the tensor contraction of this new generalized version of the Einstein equation restores the usual 2-index general relativity equation. \\
In section 5 we impose total energy-momentum conservation, and show that the generalized equation can be written with the double dual Riemann tensor $^{*}R_{ijkl}^{*}$. In the last part, the cosmological constant is also introduced.
\section{Lagrangian formulation }
We want that the new equation of general relativity necessarily contains the Riemann tensor linearly, and thus be a fourth-order tensor equation with the same index symmetries as $R_{ijkl}$. To be as general as possible, this equation must also contain fourth-order tensors constructed from the Ricci tensor $R_{jl}$ and also from the scalar curvature $R$. 
To apply the principle of least action, we must first define a very general gravitational action, $S_{(G)}$, written with the Lagrangian, $L_{(G)}$, including all these various terms linearly:   
\begin{equation}
S_{(G)}=-\frac{1}{2 \chi c}\,\int L_{(G)}( R_{ijkl}, R_{jl}, R)\sqrt{-g} \, d\Omega  
\end{equation}
where $\chi=8 \pi G/c^{4}$ is the Einstein constant. 
    \subsection{Riemann tensor symmetries}
The Riemann tensor symmetries are well known in the context of general relativity; antisymmetry on the first two indices, antisymmetry on the last two indices and symmetry obtained by exchanging the first pair with the second pair \cite{citLandau,citMisner}: 
\begin{equation}
R_{ijkl} =-R_{jikl}=-R_{ijlk}=R_{klij}\label{symRijkl}
\end{equation}
The Ricci tensor, which comes from the contraction of the Riemann tensor $ R_{jl} =  g^{ik} R_{ijkl}  $, is therefore symmetric on its two indices, and the scalar curvature $ R = g^{jl} R_{jl} $, obviously does not have any symmetry. To construct two fourth-order tensors from $ R_{jl} $ and $ R $, and having exactly the same symmetries as the Riemann tensor, we shall naturally take a combination involving metric tensors. It is easy to check that the combinations $(g_{ik}R_{jl}-g_{jk}R_{il}+g_{jl}R_{ik}-g_{il}R_{jk})$ and $(g_{ik}g_{jl}R-g_{il}g_{jk}R)$ are unique and obey exactly the same symmetries as in (\ref{symRijkl}). For convenience in what follows, we define the tensor: $g_{ijkl}=g_{ik}g_{jl}-g_{il}g_{jk}$, and we note:
\begin{align}
g_{ik}R_{jl}-g_{jk}R_{il}+g_{jl}R_{ik}-g_{il}R_{jk} &= g_{ijkp}R^{p}{}_{l}+g_{ijpl}R^{p}{}_{k}  \label{gijklRik} \\
g_{ik}g_{jl}R-g_{il}g_{jk}R &= g_{ijkl}R  \label{gijklR}
\end{align}
These combinations of metric tensors are often found in many well known references, \cite{citLandau,citMisner,citWeinberg,citEisenhart,citCaroll,citBlau} and it is therefore these specific fourth-order tensors we will use in our calculations.
\subsection{Lagrangian terms}
The Einstein-Hilbert Lagrangian of the general relativity is purely gravitational and is defined by the scalar curvature \cite{citLandau,citMisner}: 
\begin{equation}
 L_{(G)} = R  \label{lagrangien1a}
\end{equation}
However, it is well known that to obtain Einstein's familiar 2-index equation by a least action principle, we have to perform the calculations using a action including a Lagrangian written with the Ricci tensor, $L_{(G)} = g^{ik}R_{ik}$: 
\begin{align}
S_{(G)}& =  -\frac{1}{2 \chi c}\,\int R\sqrt{-g} \, d\Omega \nonumber \\
     & =-\frac{1}{2 \chi c}\,\int g^{ik}R_{ik}\sqrt{-g} \, d\Omega  \label{lagrangien1b}
\end{align}
In this paper, we remain within the familiar framework of general relativity, and so it is also physically equivalent to perform the calculations using a generalized action constructed from the three interesting tensors, (\ref{symRijkl}), (\ref{gijklRik}) and (\ref{gijklR}):
\begin{align}
 S_{(G)} = &-\frac{1}{2 \chi c}\,\int g^{ik}g^{jl}\,[ a_{1}R_{ijkl}+{a_{2}}\,(g_{ijkp}R^{p}{}_{l}  \nonumber\\ 
&+g_{ijpl}R^{p}{}_{k}) +{a_{3}}\,g_{ijkl}R\,]\sqrt{-g} \, d\Omega  \label{lagrangien1c}
\end{align}
where $a_{1}$, $a_{2}$, $a_{3}$ are three arbitrary parameters that can be determined by contraction and identification with relations (\ref{lagrangien1b}):
\begin{equation}
a_{2}=\frac{(1-a_{1})}{(n-2)} \quad \mbox{,} \quad   a_{3}=-\frac{(1-a_{1})}{(n-1)(n-2)} \, \label{a2a3} 
\end{equation}
Here, we have used the following mathematical relationships for the tensorial contractions:
\begin{align}
 g^{jl}R_{ijkl}&=R_{ik}  \label{cont1a} \\
 g^{jl}(g_{ijkp}R^{p}{}_{l}+g_{ijpl}R^{p}{}_{k})&=(n-2)R_{ik}+g_{ik}R  \label{cont1b} \\     
 g^{jl}(g_{ijkl}R)&=(n-1)g_{ik}R  \label{cont1c} \\
g^{ik}g^{jl}R_{ijkl}&=R   \label{conta}  \\
g^{ik}g^{jl}(g_{ijkp}R^{p}{}_{l}+g_{ijpl}R^{p}{}_{k})&=2(n-1)R  \label{contb}  \\     
g^{ik}g^{jl}(g_{ijkl}R)&=n(n-1)R  \label{contc}
\end{align}
with $g^{jl}g_{jl}=\delta_{j}^{j}=n$ in a \textit{n}-dimensional space.\\ 
The two relations (\ref{a2a3}) therefore, allow us to write a generalized Lagrangian in a form that involves only one parameter $a$ (with $a=a_{1}$): 
\begin{align}
L_{(G)} = & \, g^{ik}g^{jl}\,\big[\,aR_{ijkl}+\frac{(1-a)}{(n-2)}\,(g_{ijkp}R^{p}{}_{l}+g_{ijpl}R^{p}{}_{k}) \nonumber \\ 
&- \frac{(1-a)}{(n-1)(n-2)}\,g_{ijkl}R \,\big]    \label{lagrangien2} 
\end{align}
This new Lagrangian is a natural generalization, and we can verify that the contraction with $g^{ik}g^{jl}$ naturally restores the Einstein-Hilbert Lagrangian, $L_{(G)}=R$, whatever the values of $a$ and $n$. This Lagrangian is therefore physically compatible with the general theory of relativity and a 4-index Einstein equation will be obtained in the next chapter by a variational principle using the relation (\ref{lagrangien2}).  
\section{Principle of least action}
It is well known that the total action $S$ is the sum of a purely gravitational Einstein-Hilbert action $S_{(G)}$, and a matter-field action $S_{(MF)}$ which describes any matter and fields living on the space-time \cite{citLandau}:
\begin{align}
S=S_{(G)}+S_{(MF)}= & -\frac{1}{2 \chi c}\,\int L_{(G)}\sqrt{-g} \, d\Omega \nonumber 
\\                  & +\frac{1}{c}\,\int L_{(MF)}\sqrt{-g} \, d\Omega  \label{intactiong}
\end{align}
In order to obtain the gravitation field equation, we should vary the total action with respect to the metric:  
\begin{align}
 \delta S=\delta S_{(G)}+\delta S_{(MF)}= & -\frac{1}{2 \chi c}\,\delta\int L_{(G)}\sqrt{-g} \, d\Omega  \nonumber
\\ & +\frac{1}{c}\,\delta\int L_{(MF)}\sqrt{-g} \, d\Omega   \label{deltaaction}
\end{align}
The variation of $S_{(MF)}$ is given by:
\begin{align}
\delta S_{(MF)}&=\frac{1}{c}\, \delta \int L_{(MF)} \sqrt{-g} \, d\Omega \nonumber \\
             &= \frac{1}{2c}\, \int T_{jl}\, \delta g^{jl} \sqrt{-g} \, d\Omega \quad \nonumber \\
						&= \frac{1}{2c}\, \int  \, T_{ijkl} \, g^{ik} \delta g^{jl} \, \sqrt{-g} \, d\Omega  \label{actione2}
\end{align}
where $T_{jl}$ is the familiar total energy-momentum tensor defined in \cite{citLandau}: 
\begin{align}
T_{jl}=\frac{2}{\sqrt{-g}}  \, \frac{ \delta \, ( \, L_{(MF)} \,\sqrt{-g} \,) }{ \delta g^{jl}} 
\end{align}
, and $T_{ijkl}$ a generalized 4-index total energy-momentum tensor such that:
\begin{align}
T_{jl}= g^{ik}\,T_{ijkl}
\end{align}
The variation of $S_{(G)}$ is obtained by using the lagrangian (\ref{lagrangien2}):
\begin{align}
\delta S_{G}=&-\frac{1}{2 \chi c}\,\delta \int L_{(G)} \sqrt{-g}\, d\Omega  \nonumber  \\
=&-\frac{1}{2 \chi c}\,\delta \int  g^{ik}g^{jl} \, \big[\,aR_{ijkl}+\frac{(1-a)}{(n-2)}(g_{ijkp}R^{p}{}_{l}  \nonumber \\
 & +g_{ijpl}R^{p}{}_{k}) -\frac{(1-a)}{(n-1)(n-2)} \, g_{ijkl}R \, \big] \sqrt{-g}\,d\Omega    \nonumber \\ 
=& -\frac{1}{2 \chi c}\,\int \bigg[ \, g^{ik}g^{jl} \, \big[\, a \, \delta R_{ijkl}+ \frac{(1-a)}{(n-2)}\,\delta (g_{ijkp}R^{p}{}_{l}  \nonumber \\
& +g_{ijpl}R^{p}{}_{k}) - \,\frac{(1-a)}{(n-1)(n-2)} \,  \delta (g_{ijkl}R)\big]\sqrt{-g}   \nonumber \\ 
 & + \big[aR_{ijkl}+\frac{(1-a)}{(n-2)}\,(g_{ijkp}R^{p}{}_{l}+g_{ijpl}R^{p}{}_{k})   \nonumber\\ 
& -\frac{(1-a)}{(n-1)(n-2)}\,g_{ijkl}R\big] \, \big[ \, 2\,g^{ik}\delta g^{jl}\sqrt{-g}   \nonumber \\ 
& +g^{ik}g^{jl} \delta \sqrt{-g}  \, \big] \,  \bigg]d\Omega  \label{action} 
\end{align}
It is not so easy to directly calculate the variations of the different terms in the last line, and so we first demonstrate some useful mathematical relations:\\
$g^{jl}\delta R_{ijkl}$ using (\ref{cont1a}): 
\begin{align}
 g^{jl}\delta R_{ijkl} &= \delta    \big[\,g^{jl}R_{ijkl}\,\big]-\delta g^{jl}R_{ijkl} \nonumber \\      
								       &=\delta R_{ik}-\delta g^{jl}R_{ijkl}  \label{deltaRijkl1} 
\end{align} 
$g^{jl}\delta (g_{ijkp}R^{p}{}_{l}+g_{ijpl}R^{p}{}_{k})$ using (\ref{cont1b}):
\begin{align}
g^{jl}\delta (g_{ijkp}R^{p}{}_{l}+g_{ijpl}R^{p}{}_{k})&=\delta \big[g^{jl} (g_{ijkp}R^{p}{}_{l}+g_{ijpl}R^{p}{}_{k})\big] \nonumber \\
&-\delta g^{jl}(g_{ijkp}R^{p}{}_{l}+g_{ijpl}R^{p}{}_{k})  \nonumber \\
                                             &=(n-2)\delta R_{ik} + \delta(g_{ik}R)  \nonumber \\
																						&-\delta g^{jl}(g_{ijkp}R^{p}{}_{l}+g_{ijpl}R^{p}{}_{k})    \label{deltaRijkl2} 
\end{align}
$g^{jl}\delta (g_{ijkl}R)$ using (\ref{cont1c}): 
\begin{align}
g^{jl} \delta (g_{ijkl}R)&=\delta \big[\, g^{jl}(g_{ijkl}R)\,\big]-\delta g^{jl}(g_{ijkl}R)  \nonumber \\
                         &= (n-1)\delta(g_{ik}R)-\delta g^{jl}(g_{ijkl}R)  \label{deltaRijkl3}
\end{align}
\\
$ \delta \sqrt{-g}=-1/2 \,g_{jl}\, \delta g^{jl}\,\sqrt{-g} $ \cite{citLandau}, and using \\ $g_{ijkl}\,g^{ik}=(n-1)g_{jl}$ we can write: 
\begin{equation}
\delta \sqrt{-g}=-\frac{1}{2(n-1)}\,g_{ijkl}\,g^{ik} \delta g^{jl}\,\sqrt{-g}  \label{deltaRijkl4} 
\end{equation}
The relationships (\ref{deltaRijkl1}), (\ref{deltaRijkl2}), (\ref{deltaRijkl3}), (\ref{deltaRijkl4}) allow one to rewrite $\delta S_{(G)}$ (\ref{action}) in the form:
\begin{align}
\delta S_{(G)}=& -\frac{1}{2 \chi c}\,\int \big[\,aR_{ijkl}+ \frac{(1-a)}{(n-2)}\,(g_{ijkp}R^{p}{}_{l}+g_{ijpl}R^{p}{}_{k}) \nonumber  \\  
  &-\frac{(n-2a)}{2(n-1)(n-2)} \, g_{ijkl}R\,\big]g^{ik}\delta g^{jl}\sqrt{-g}  \, d\Omega \nonumber \\
  & +\int g^{ik}\delta R_{ik}  \sqrt{-g} \, d\Omega \label{action2} 
\end{align}
We note that the terms containing $\delta(g_{ik}R)$ have been simplified by themselves. The terms containing $g^{ik}\delta R_{ik}$ result in an integral of a covariant divergence, and hence by Stoke's theorem are equal to a boundary contribution at infinity which we can set to zero by making the variation vanish at infinity  \cite{citCaroll,citBlau}: $\int g^{ik}\delta R_{ik} \, \sqrt{-g} \, d\Omega=0$. \\
Using the relations (\ref{actione2}) and (\ref{action2}), the variation of the total action, $\delta S$, thus becomes: 
\begin{align}
\delta S&=\delta S_{(G)}+\delta S_{(MF)} \nonumber \\ 
&= -\frac{1}{2\chi c} \int \big[aR_{ijkl}+ \frac{(1-a)}{(n-2)}(g_{ijkp}R^{p}{}_{l}+g_{ijpl}R^{p}{}_{k})  \nonumber \\ 
 &-\frac{(n-2a)}{2(n-1)(n-2)} \, g_{ijkl}R-\chi T_{ijkl} \big]g^{ik}\delta g^{jl}\sqrt{-g} \, d\Omega  \label{actiontot}
\end{align}
\section{Generalized 4-index Einstein equation}
    \subsection{Fourth-order equation}
Since the expression for the variation of the total action $\delta S$ (\ref{actiontot}) should hold for any variation $ \delta g^{jl}$, we must have:  
\begin{multline}
g^{ik}\big[aR_{ijkl}+ \frac{(1-a)}{(n-2)}(g_{ijkp}R^{p}{}_{l}+g_{ijpl}R^{p}{}_{k}) \\ - \frac{(n-2a)}{2(n-1)(n-2)} g_{ijkl}R-\chi T_{ijkl}\big]= 0  \label{eq1}
\end{multline}
In general, for arbitrary $T_{ijkl}$, the expression in the parenthesis is not equal to zero, and we can not simply exclude the contractional factor $g^{ik}$. However, it is always mathematically possible to choose a particular energy-momentum tensor $T_{ijkl}$ which has the same number of components and the same symmetries as $R_{ijkl}$, and which acts as a source term, allowing us to write the sought after generalized 4-index Einstein equation:
\begin{multline}
aR_{ijkl}+ \frac{(1-a)}{(n-2)}(g_{ijkp}R^{p}{}_{l}+g_{ijpl}R^{p}{}_{k}) \\ - \frac{(n-2a)}{2(n-1)(n-2)} \, g_{ijkl}R =\chi \, T_{ijkl}  \label{eqgene2}
\end{multline}
We can easily check that the result of the tensorial contraction of this equation gives the famous Einstein equation of general relativity, whatever the values of $a$ and $n$ (as surprising as it may seem): 
\begin{equation}
R_{jl}-\frac{1}{2}\,g_{jl}R=\chi \, T_{jl}  \label{eqeinstein}
\end{equation}
This is of course an important result for checking our calculations, pointing out that we did not use Einstein's equation in any intermediate calculations to obtain the generalized equation (\ref{eqgene2}). This first result also shows us unambiguously that the 2-index Einstein equation (\ref{eqeinstein}) has the same form in any space-time of dimension $n$, which is also an important result.\\
To our knowledge, this new 4-index gravitational field equation, obtained by means of a rigorous mathematical treatment using the variational principle, is written here for the first time. Due to the presence of the Riemann tensor, and therefore the Weyl tensor, we will show later that equation (\ref{eqgene2}) contains, as expected, more information than equation (\ref{eqeinstein}).  
	  \subsection{Fourth-order Einstein tensor}
The final equation (\ref{eqgene2}) can also be written in a more compact form: 
\begin{equation}
G_{ijkl}=\chi\,T_{ijkl}  \label{eq2} 
\end{equation}
where we introduced here a new fourth-order tensor:
\begin{align}
G_{ijkl}=& \, aR_{ijkl}+ \frac{(1-a)}{(n-2)}\,(g_{ijkp}R^{p}{}_{l}+g_{ijpl}R^{p}{}_{k}) \nonumber \\ 
&- \frac{(n-2a)}{2(n-1)(n-2)} \, g_{ijkl}R  \label{Gijkl1}
\end{align}
We can verify that:
\begin{equation}
 g^{ik}G_{ijkl}= R_{jl}-\frac{1}{2}\,g_{jl}R=G_{jl}
\end{equation}
where we recognize here the famous 2-index Einstein tensor $G_{jl}$. The fourth-order tensor $G_{ijkl}$ containing the Riemann tensor, the Ricci tensor and the scalar curvature, can therefore be considered as a generalization of $G_{jl}$. 
	  \subsection{Weyl tensor}
The fourth-order Einstein tensor $G_{ijkl}$, depends partly on the mathematical parameter $a$ and can therefore be split in two distinct parts of which only one depends on $a$:
\begin{equation}
G_{ijkl}=aC_{ijkl} \,+\, B_{ijkl}    \label{GijklCijkl1} 
\end{equation} 
and we can write by identification with (\ref{Gijkl1}):
\begin{align}
 C_{ijkl} = & \,R_{ijkl}-\frac{1}{(n-2)}(g_{ijkp}R^{p}{}_{l}+g_{ijpl}R^{p}{}_{k}) \nonumber \\
&+\frac{1}{(n-1)(n-2)}g_{ijkl}R     \label{Cijklb} 
\end{align}  
\begin{align}
 B_{ijkl}= & \, \frac{1}{(n-2)}\,(g_{ijkp}R^{p}{}_{l}+g_{ijpl}R^{p}{}_{k}) \nonumber \\
&-\frac{n}{2(n-1)(n-2)} \, g_{ijkl}R  \label{Bijkl}
\end{align}  
We recognize here the Weyl tensor $C_{ijkl}$ written in $n$ dimensions \cite{citWeinberg,citEisenhart,citCaroll,citBlau,citOyv}, and a new, hitherto unknown tensor $B_{ijkl}$. It is easy to check that the tensorial contractions of (\ref{Cijklb}) and (\ref{Bijkl}) give: 
\begin{align}
 g^{ik}C_{ijkl} &= 0   \label{gikCijkl} \\
 g^{ik}B_{ijkl}&= R_{jl}-\frac{1}{2}\,g_{jl}R \label{gikBijkl} 
\end{align}  
This last result clearly show us that some of the information contained in the generalized equation (\ref{eq2}), is not recovered in the 2-index Einstein equation (\ref{eqeinstein}). This is the part about the Weyl tensor.
	  \subsection{Two part decomposition of total energy-momentum tensor $T_{ijkl}$}
It is well known that Einstein himself and a major consensus of famous physicists (see \cite{citLandau,citWeinberg,citSchro,citDirac} for example) have emphasized that the gravitational field must also have an energy-momentum tensor, as do all other physical fields. In this paper, the 4-index tensor $T_{ijkl}$, which describes any matter present plus the gravitational field contained in space-time, can be simply and naturally divided into two parts:
\begin{equation}
T_{ijkl}=\,T^{(M)}_{ijkl} \,+\,T^{(F)}_{ijkl}    \label{TFMijkl}
\end{equation}  
where $T^{(M)}_{ijkl}$ represents the generalized energy momentum tensor of the matter content, and $T^{(F)}_{ijkl}$ the energy-momentum tensor of the gravitational field itself \cite{citZakir}. With (\ref{GijklCijkl1}) and (\ref{TFMijkl})  the generalized equation (\ref{eq2}) becomes:
\begin{align}
 %G_{ijkl}&=\chi\,T_{ijkl} \nonumber \\
aC_{ijkl} \,+\, B_{ijkl} &= \chi \,(\,T^{(F)}_{ijkl} \,+\,T^{(M)}_{ijkl}\,) \label{GijklCijkl} 
\end{align}
		\subsection{Energy-momentum tensor of the matter $T^{(M)}_{ijkl}$ }
By rearranging the Einstein equation (\ref{eqeinstein}), it is possible to rewrite the Ricci tensor $R_{jl}$ and the scalar curvature $R$ in the form:    
\begin{align}
&R_{jl}=\chi \, (T_{jl}-\frac{1}{(n-2)}\,g_{jl}T) \label{Rjl}\\
&R=-\frac{2}{(n-2)}\,\chi \,T \label{R}
\end{align}
which we now substitute into (\ref{Bijkl}), obtaining:
\begin{align}
B_{ijkl}=&\,\chi \,\big[\,\frac{1}{(n-2)} \, (g_{ijkp}T^{p}{}_{l}+g_{ijpl}T^{p}{}_{k}) \nonumber \\ 
&-\frac{1}{(n-1)(n-2)}\,g_{ijkl}T\,\big]   \label{Tijklm}
\end{align} 
Equations (\ref{gikBijkl}) and (\ref{Tijklm}) together indicate that the tensor $B_{ijkl}$ is directly linked to the energy-momentum tensor of the matter content present in the standard theory of general relativity by means of the tensor $T_{jl}$. The right-hand side of (\ref{Tijklm}) represents a 4-index energy-momentum tensor for matter, and in our theory there is only one tensor that can play this role, namely $T^{(M)}_{ijkl}$:
\begin{align}
T^{(M)}_{ijkl} =&\,\frac{1}{(n-2)} \, (g_{ijkp}T^{p}{}_{l}+g_{ijpl}T^{p}{}_{k}) \nonumber \\
&-\frac{1}{(n-1)(n-2)}\,g_{ijkl}T
\end{align}
which allows us to write:
\begin{equation}  
B_{ijkl}=\chi \,T^{(M)}_{ijkl} \label{decTijklb}
\end{equation}
The energy-momentum tensor for matter $T^{(M)}_{ijkl}$ acts as a source term for $B_{ijkl}$, and it is not difficult to check that: 
\begin{align}
& g^{ik}T^{(M)}_{ijkl}=T_{jl}  \nonumber \\
& g^{jl}T^{(M)}_{ijkl}=T_{ik} \nonumber \\
& T^{(M)}_{ijkl}=-T^{(M)}_{jikl}=-T^{(M)}_{ijlk}=T^{(M)}_{klij} 
\end{align} 
In addition, one can easily check that the tensorial contraction of equation (\ref{decTijklb}) gives Einstein's equation (\ref{eqeinstein}).
		\subsection{Energy-momentum tensor of the gravitational field $T^{(F)}_{ijkl}$} 
With equation (\ref{GijklCijkl}), simplified by (\ref{decTijklb}), we find a simple connection between the energy-momentum tensor for the free gravitational field $T^{(F)}_{ijkl}$, and the Weyl tensor $C_{ijkl}$: 
\begin{equation}
aC_{ijkl} = \chi \,T^{(F)}_{ijkl} \label{decTijkla} 
\end{equation} 
$T^{(F)}_{ijkl}$ acts as a source term for the Weyl tensor, and the properties of $C_{ijkl}$ imply that:
\begin{align}
& g^{ik}T^{(F)}_{ijkl}=0 \nonumber \\
& g^{jl}T^{(F)}_{ijkl}=0  \nonumber \\
& T^{(F)}_{ijkl}=-T^{(F)}_{jikl}=-T^{(F)}_{ijlk}=T^{(F)}_{klij}
\end{align} 
The tensorial contraction of equation (\ref{decTijkla}) is equal to zero. The two tensors, $C_{ijkl}$ and $T^{(F)}_{ijkl}$, are logically linked because they only concern the free gravitational field in vacuum, and it is precisely this part which is not present in the standard theory of general relativity. The information contained in equation (\ref{decTijkla}) is not contained in the 2-index Einstein equation, and for this reason a fourth-order theory is fully justified. \\
For a vanishing Ricci tensor, $R_{jl}=0$, the Riemann curvature tensor is equal to the Weyl tensor, $R_{ijkl}=C_{ijkl}$, and it is well known that the latter describes the free gravitational field in vacuum and provides curvature to space-time \cite{citBlau,citOyv}. A flat space-time implies no matter with $T^{(M)}_{ijkl}=0$, no field with $T^{(F)}_{ijkl}=0$, and likewise $R_{ijkl}=0$, as can be seen from the generalized Einstein equation (\ref{eqgene2}).\\
The total energy-momentum tensor, $T_{ijkl}=T^{(M)}_{ijkl}+T^{(F)}_{ijkl}$, gives the most complete description of the medium with respect to its effect on the geometry. It describes both the state of matter and the state of the gravitational field in vacuum, but it appears to be very non-trivial to find the form of the tensor $T^{(F)}_{ijkl}$. The determination of this tensor, however, and therefore the form of equation (\ref{decTijkla}), would allow us to find important solutions concerning the contribution of the gravitational field itself. The tensor $T^{(F)}_{ijkl}$ naturally takes its place in the new generalized 4-index equation, and we'll show in the next chapter that $T_{ijkl}$ obeys total energy-momentum conservation.    
		\subsection{Energy-momentum conservation}
In the preceding chapters the parameter, $a$, has been treated as a mathematical parameter with no particular meaning. We will now show that $a$ takes on a particular physical value in the context of covariant conservation of the total 4-index energy-momentum tensor:  
\begin{equation}
\nabla_{i}T^{i}{}_{jkl}=\nabla_{i}(T^{(F)}{}^{i}{}_{jkl}+T^{(M)}{}^{i}{}_{jkl})=0  \label{Tijklan2}
\end{equation} 
where $\nabla _{i}$ is the covariant derivative operator.\\
With the definition of the generalized Enstein tensor (\ref{Gijkl1}), and using the usual metric compatibility $\nabla _{i} \, g_{jl}=0$, we can write:
\begin{align}
\nabla _{i} G^{i}{}_{jkl}
=& \nabla _{i} \big[aR^{i}{}_{jkl}+ \frac{(1-a)}{(n-2)}(\delta^{i}_{k} R_{jl} - g_{jk}R^{i}{}_{l} + g_{jl}R^{i}{}_{k} \nonumber \\
& -\delta^{i}_{l}R_{jk}) - \frac{(n-2a)}{2(n-1)(n-2)} (\delta^{i}_{k} g_{jl}-\delta^{i}_{l} g_{jk} )R \big] \nonumber \\
= &\, a\nabla _{i}R^{i}{}_{jkl}+ \frac{(1-a)}{(n-2)}\,(\nabla _{k} R_{jl} - g_{jk}\nabla _{i} R^{i}{}_{l} \nonumber \\  
&+ g_{jl}\nabla _{i}R^{i}{}_{k} - \nabla _{l} R_{jk})  \nonumber \\
 &- \,\frac{(n-2a)}{2(n-1)(n-2)} \,(g_{jl}\nabla _{k} R-g_{jk}\nabla _{l} R )   \nonumber 
\end{align}
We make use of the contracted Bianchi identities, \cite{citWeinberg,citCaroll}: $\nabla _{i}R^{i}{}_{jkl}=\nabla _{k} R_{jl}-\nabla _{l} R_{jk}$ and  $\nabla _{i} R^{i}{}_{k}=1/2 \,\nabla _{k} R$ to deduce: 
\begin{align}
\nabla _{i} G^{i}{}_{jkl}= &\, - \frac{1+a(n-3)}{(n-2)} \big[\nabla _{l} R_{jk} - \nabla _{k} R_{jl} \nonumber \\
& + \frac{1}{2(n-1)} (g_{jl}\nabla _{k} R-g_{jk}\nabla _{l} R) \big]  \nonumber \\
=&\, - \frac{1+a(n-3)}{(n-2)} \, C_{jkl} \label{nablaGijkl1}
\end{align}
where we identify here the Cotton tensor $C_{jkl}$ \cite{citCotton,citCotton2}.\\
Imposing energy conservation thus gives us the physical value of $a$:
\begin{equation}
\nabla _{i} G^{i}{}_{jkl}=0  \,\, \Rightarrow \,\,    \nabla_{i}T^{i}{}_{jkl}=0 \,\,  \Rightarrow \,\, a=- \frac{1}{(n-3)} \label{Tijklan3}
\end{equation} 
       \section{Generalized equation with $a=-1/(n-3)$}
\subsection{Double-dual Riemann tensor $^{*}R_{ijkl}^{*}$ }
In the case where energy is conserved, taking $a=-1/(n-3)$, the generalized 4-index Einstein equation of general relativity (\ref{eqgene2}) becomes:
\begin{equation}
\frac{1}{(n-3)}\,\big[-R_{ijkl}+ g_{ijkp}R^{p}{}_{l}+g_{ijpl}R^{p}{}_{k} - \frac{1}{2}  g_{ijkl}R  \big] =\chi  T_{ijkl}  \label{DDrijkl3}
\end{equation} 
where we recognize here the double (Hodge) dual Riemann tensor \cite{citMisner,citGliner}:
\begin{align}
^{*}R_{ijkl}^{*}= &\frac{1}{4}\, e_{ijpq}\, R^{pqrs}e_{klrs}  \nonumber \\
=&-R_{ijkl}+ g_{ijkp}R^{p}{}_{l} +g_{ijpl}R^{p}{}_{k} - \frac{1}{2} \, g_{ijkl}R \label{DDrijkl1}
\end{align}
with $e_{ijkl}$ the Levi-Civita tensor. \\
The general equation (\ref{DDrijkl3}) can thus be written simply as:
\begin{equation}
^{*}R_{ijkl}^{*}= \chi \, (n-3) \, T_{ijkl} \label{DDrijkl3b} 
\end{equation} 
It is not difficult to check that the tensorial contraction of this equation again yields the Einstein equation (\ref{eqeinstein}), with $g^{ik} {} ^{*}R_{ijkl}^{*}= (n-3) (R_{jl}-\frac{1}{2}\,g_{jl}R)$,  and that (\ref{DDrijkl3b}), as expected, also satisfies total energy-momentum conservation:
\begin{align}
&\nabla _{i}{}^{*}R^{*}{}^{i}{}_{jkl}=0  \\
&\nabla _{i}T^{i}{}_{jkl}=\nabla _{i}(T^{(F)}{}^{i}{}_{jkl}+T^{(M)}{}^{i}{}_{jkl})=0   
\end{align} 
According to (\ref{decTijklb}) and (\ref{decTijkla}), it is also possible to write equation (\ref{DDrijkl3b}) into two parts:
\begin{align}
C_{ijkl} &= -\chi  (n-3) \, T^{(F)}_{ijkl}  \label{eqfinaleb} \\
B_{ijkl} &=   \chi \, T^{(M)}_{ijkl}  \quad \label{eqfinalea}
\end{align}
As we have already seen, these two equations are not coupled; they are independent, and their solutions found individually must be added to satisfy total energy momentum conservation. As an analogy in classical mechanics, for example, it is akin to the kinetic energy and the potential energy that derive from different relations, but which must finally be added to have conservation of the total mechanical energy.
\subsection{Particular solution of the equation  $B_{ijkl}=\chi \, T^{(M)}_{ijkl}$}
In the very important case of a centrally symmetric field in vacuum, that is, outside of the masses producing the field, $T^{(M)}_{ijkl}=0$, it is possible to solve the equation (\ref{eqfinalea}), for $n=4$:
\begin{equation}
B_{ijkl}{}_{(n=4)}=  \frac{1}{2}\,(g_{ijkp}R^{p}{}_{l}+g_{ijpl}R^{p}{}_{k}) - \,\frac{1}{3} \, g_{ijkl}R=0 \label{Bijklvac}
\end{equation}  
If we use spherical space coordinates $(r,\theta,\phi)$, then the general expression for $ds^2$ is:
\begin{equation}
ds^2= A(r)c^2dt^2-B(r)dr^2-r^2d\theta^2-r^2sin^2\theta d\phi^2
\end{equation} 
To get the differential equations of gravitation with the functions $A(r)$ and $B(r)$, we must first calculate the components of the tensor $B_{ijkl}{}_{(n=4)}$. Using a familiar method, it is not difficult (but long) to obtain the following independent relations:
\begin{align}
B_{0101} =  & \,-\frac{1}{6}\, \big[\,A^{''}-\frac{A^{'2}}{2A} -\frac{A^{'}B^{'}}{2B}+\frac{AB^{'}}{rB} \nonumber \\ 
 & + \frac{4AB}{r^2} -\frac{A^{'}}{r}-\frac{4A}{r^2} \,  \big]   \\
B_{0202} = &\, \frac{r^2}{12B}\, \big[\,A^{''}-\frac{A^{'2}}{2A} -\frac{A^{'}B^{'}}{2B} -\frac{5AB^{'}}{rB} \nonumber \\ 
& - \frac{2AB}{r^2} -\frac{A^{'}}{r}+\frac{2A}{r^2} \,  \big]  \\
B_{1212} = &\, -\frac{r^2}{12A}\, \big[\,A^{''}-\frac{A^{'2}}{2A} -\frac{A^{'}B^{'}}{2B}  +\frac{AB^{'}}{rB} \nonumber \\ 
&  - \frac{2AB}{r^2} +\frac{5A^{'}}{r}+\frac{2A}{r^2} \,  \big]  \\
B_{2323} = &\, -\frac{r^{4} sin^2\theta}{3AB}\, \big[\,A^{''}-\frac{A^{'2}}{2A} -\frac{A^{'}B^{'}}{2B}  -\frac{AB^{'}}{2rB} \nonumber \\ 
&  +\frac{AB}{r^2} +\frac{A^{'}}{2r}-\frac{A}{r^2} \,  \big]  
\end{align}
By combining these differents equations, we finally obtain a well known solution, the standard Schwarzschild metric for the space-time \cite{citLandau,citMisner}: 
\begin{align}
 A(r)=& \, \frac{1}{B(r)}=1+\frac{r_g}{r}  \\
 ds^2=& \, (1+\frac{r_g}{r})\,c^2dt^2-(1+\frac{r_g}{r})^{-1}dr^2 \nonumber \\ 
 &-r^2d\theta^2-r^2sin^2\theta d\phi^2
\end{align} 
This is not surprising because equation (\ref{Bijklvac}) contains only the Ricci tensor and the scalar curvature, and therefore we expect to find the same solutions as those obtained with the Einstein equation (a good sign). Equation (\ref{eqfinalea}), however, goes much further because it may allow us to find the exact solutions for any dimension of space-time $n$. \\
Because the generalized equation (\ref{DDrijkl3b}) is contracted to obtain the original Einstein equation, only standard solutions are present in the familiar 2-index general theory of relativity, but what is lacking are the solutions of equation (\ref{eqfinaleb}) giving the contribution of the gravitational field itself in vaccum. The next step in our future research will be to determine precisely the mathematical composition of the energy-momentum tensor $T^{(F)}_{ijkl}$ in order to obtain solutions to equation (\ref{eqfinaleb}), but this appears to be a very difficult task.  
\subsection{Cosmological constant}
It is interesting to verify that the equation (\ref{DDrijkl3b}) can be deduced from a principle of least action using the Lagrangian (\ref{lagrangien2}) written with $a=-1/(n-3)$:
\begin{align}
L_{(G)} = & \,\frac{1}{(n-3)} \, g^{ik}g^{jl}\,\big[-R_{ijkl}+ g_{ijkp}R^{p}{}_{l}+g_{ijpl}R^{p}{}_{k} \nonumber \\
&-\frac{1}{(n-1)}\,g_{ijkl}R \, \big]     \label{lagrangien3}
\end{align}
We will work again within the framework of general relativity, where the cosmological constant $\Lambda $ is introduced into the Lagrangian by an additional term \cite{citCaroll}:  $ -2\Lambda $. In the same way as in (\ref{contc}), this term can be rewritten as:   $ -2\Lambda= -2/{n(n-1)} \, g^{ik}g^{jl}\,(g_{ijkl}\,\Lambda)$. The principle of least action applied to this term modifies equation (\ref{DDrijkl3b}), giving us:  
\begin{equation}
^{*}R_{ijkl}^{*}+\frac{(n-3)}{(n-1)}\,g_{ijkl}\,\Lambda= \chi \, (n-3) \,T_{ijkl}   \label{DDrijklcosmo}
\end{equation}
%
% and equation (\ref{eq2}) becomes: 
%
%\begin{equation}
%G_{ijkl}+\frac{1}{(n-1)}\,g_{ijkl}\,\Lambda=\chi \, T_{ijkl} \label{eqgenecosmo}
%\end{equation} 
%
Contracting this latter equation indeed restores Einstein's standard equations with the cosmological constant \cite{citCaroll,citBlau}:
\begin{equation}
R_{jl}-\frac{1}{2}\,g_{jl}R + g_{jl}\Lambda =\chi \, T_{jl}  \label{eqeinsteinconst}
\end{equation}
\section{Conclusion}
In this paper we have rigorously demonstrated by using the principle of least action, that there exists a much more general equation than that of Einstein. This new 4-index equation explicitly and linearly includes the Riemann tensor and hence the Weyl tensor as well as the energy of the gravitational field $T^{(F)}_{ijkl}$. The contraction of this equation restores the usual 2-index Einstein equation. The paper highlights two main results: \\
- the first result is given by the equation, $B_{ijkl}=\chi\,T^{(M)}_{ijkl}$,  which could allow us to find the exact Schwarzschild solutions for any $n$ dimension space. \\
- the second important result is given by the equation, $C_{ijkl} =-\chi(n-3)\,T^{(F)}_{ijkl}$, containing the Weyl tensor and the energy-momentum tensor of the gravitational field in vaccum, which could be connected, as we are hoping, to the mysterious dark energy. \\
The next step in our research will be to determine precisely the composition of the energy-momentum tensor $T^{(F)}_{ijkl}$ in order to obtain more general solutions including the contribution of the gravitational field energy itself. \\

\end{document}